\documentstyle[12pt,epsfig]{article}

\def\req#1{(\ref{#1})}

\def\bra{\langle}
\def\ket{\rangle}

\def\noi{{\noindent}}

\def\nn{\nonumber \\}
\def\bc{\begin{center}}
\def\ec{\end{center}}
\def\1#1{{\bf #1}}
\def\lp{\left(}
\def\rp{\right)}

\let\a=\alpha  \let\g=\gamma 
   
   \let\m=\mu
\let\n=\nu  \let\p=\pi \let\r=\rho

   \let\Th=\Theta
  
\begin{document}

\begin{center}
{\large \bf The Van-der-Waals Gas EOS for the Lorentz Contracted Rigid  Spheres
}

\vspace{1.0cm}

{\bf Kyrill A. Bugaev
}\\

\vspace{1.cm}

\vspace{0.5cm}

Bogolyubov Institute for Theoretical Physics,\\
03680 -- Kiev, Ukraine\\

\date{\today}

\end{center}

\vspace{2.0cm}

\begin{abstract}
The relativistic  equation of state (EOS) of the Van-der-Waals gas is suggested and analyzed.
In contrast to the usual case,  the Lorentz contraction of the sphere's volume
is taken into account. It is proven that the suggested EOS 
obeys the causality
in the limit of high densities, i.e., the value of sound velocity of such a media is 
subluminar.
The pressure obtained for the high values of  chemical potential has an 
interesting kinetic interpretation. 
The suggested EOS shows that for high densities the most probable configuration
corresponds to the smallest value of the relativistic excluded volume. 
In other words, for high densities  the configurations with the collinear velocities of the neighboring 
hard core particles are the most probable ones. This, perhaps, may shed light 
on the coalescence process of any relativistic hard core constituents.

\end{abstract}

\vspace{4.cm}


\newpage

\bc
{\large \bf 1. Introduction }

\vspace{0.2cm}

\ec

The van der Waals (VdW) excluded volume model is successfully  used
to describe the hadron yields measured  in relativistic nucleus--nucleus
collisions (see e.g. \cite{Yen:99,St:99} and references therein).
This model treats the hadrons as hard core spheres and, therefore,
takes into account the hadron
repulsion at short distances. In a relativistic situation one should,
however,
include the Lorentz contraction of the hard core hadrons.
Recently, both the conventional cluster and the virial expansions
were generalized to the  momentum dependent
inter-particle potentials,
accounting for the Lorentz contracted hard core
repulsion \cite{Bug:00} and the derived equation of state  (EOS)  was applied  to describe 
hadron yields observed  in relativistic nuclear collisions \cite{Zeeb:02}.
The VdW equation obtained in the traditional way 
leads to the reduction of the second virial coefficient (analog of the
excluded volume)
compared to  nonrelativistic case.
However, in the high pressure limit the second virial coefficient
remains finite. This fact immediately leads to the problem
with causality in relativistic mechanics - the
speed of sound exceeds the  speed of light \cite{Raju:92}.

The influence of   relativistic effects on the hard core repulsion may be important for 
a variety of  effective models of hadrons  and hadronic matter
such as the  modified  
Walecka model \cite{Ris:91}, various extensions of the 
Nambu--Jona-Lasinio model \cite{Na:61}, the quark-meson coupling model  \cite{Gu:X}, 
the chiral SU(3) model \cite{chirm} e.t.c.
Clearly,  the relativistic hard core repulsion should be  important  for
any effective model in which the strongly interacting particles 
have  the reduced values of  masses compared to their vacuum values because 
with lighter masses 
the large portion of  particles  becomes  relativistic. 
Nevertheless, the relativistic hard core repulsion was, so far,  
not incorporated into these models  due to the absence of  the required formalism. 

The  Lorentz contraction  of  rigid spheres representing the hadrons  may also  be   essential  at high particle densities  
which can be achieved at modern colliders.
Very recently it was understood that in the  baryonless deconfined phase  
above the  cross-over temperature $T_c$  some hadrons 
may survive up to  large temperatures like $ 3 T_c$ 
\cite{Karsch:02, Hatsuda:03, Shur:04b},  and   that above $T_c$ 
there may exist bound states \cite{Shur:04} and resonances \cite{Rapp:05}. 
Moreover,   an exactly solvable statistical model of quark-gluon bags with surface 
tension \cite{Bugaev:07} indicates that above the cross-over transition \cite{Shur:04b} the coexistence  of hadronic resonances with QGP  may, in principle,  survive  up  to infinite temperature. 
Thus, above  $T_c$   the relativistic effects of the hard core repulsion can be important for many hardonic resonances and hadron-like  bound states  of quarks, especially, if their masses are
reduced due to chiral symmetry restoration.

%%%These and other observations led to the  conclusion that at the initial  temperatures 
%%%$T \approx  (1-2) T_c$ achieved  at RHIC quark gluon plasma is in a strongly coupled 
%%%regime (sQCD) \cite{Shur:04b}.

%%At the moment there are  a few  guesses on how to develop  the 
%%%statistical mechanics of  QGP coexisting with the hadronic resonances above 
%%%the cross-over temperature, but it is possible that the relevant 
%%%quasiparticle degrees of freedom may   include the dressed constituent 
%%%quarks or/and their  hadron-like  bound states. 
%%%In this case one can think about the possibility to describe  such states 
%%%in terms of relativistic particles with the  hard core repulsion on short distances. 
%%%In fact, according  to Shuryak \cite{Shur:05} the recent study of the strongly coupled 
%%%colored classical plasma does include the short range repulsion 
%%%of the  inverse distance square type. 

%%%Also it is quite possible that the large  values of the partonic  cross-sections which are 
%%%necessary to reproduce the  values of elliptic flow observed at RHIC energies 
%%%\cite{GyulassyMolnar} do evidence that during the course of heavy ion collision  
%%%the partons are  acquiring   some  finite transversal size.

Also 
the VdW EOS, 
which obeys the causality condition in the limit of high density 
and simultaneously reproduces the correct low density behavior,
adds a significant theoretical value 
%%%for the  relativity
because %%%,due to some unclear reasons,
such an EOS had not yet been formulated during more than a century of the special relativity.
This work is devoted to
the investigation of the necessary  assumptions  to formulate   such an equation of state. 

The work  is organized as follows.
In Sect. 2 a summary of both the cluster and  virial expansion  
for the Lorentz contracted rigid spheres is given.
It is shown that the  VdW extrapolation in relativistic case is not 
a unique procedure. 
Therefore,  an alternative derivation of  
the VdW EOS is considered there.
The high pressure limit is studied in details  in Sect. 3.
It is shown that the suggested relativistic generalization of  the earlier approach \cite{Bug:00}
obeys the causality condition.
The conclusions are given in the last section.

\vspace*{0.4cm}

\bc
{\large \bf 2. Relativization of the van der Waals EOS }
\ec

%%% TWO

\vspace*{0.2cm}

The excluded volume effect
accounts for the blocked volume
of two spheres when they touch each other.
If hard sphere particles move with relativistic velocities
it is necessary to include their Lorentz contraction
in the rest frame of the  medium.
The model suggested in Ref. \cite{Z:95}
is not satisfactory: the second virial coefficient 
$a_2=4\,v_{\rm o}$ of the VdW excluded volume model
is confused there
with the proper volume $v_{\rm o}$ of an individual particle --
the contraction effect is introduced for
the proper volume of each particle.
In order to get the correct result it is necessary to
account for the excluded volume of
two Lorentz contracted spheres.

Let ${\bf r}_i$ and ${\bf r}_j$
be the coordinates of the $i$-th and $j$-th Boltzmann particle, respectively,
and ${\bf k}_i$ and ${\bf k}_j$ be their momenta,
${\bf {\hat r}}_{ij}$ denotes
the unit vector
$ {\bf {\hat r}}_{ij} = {\bf r}_{ij}/|{\bf r}_{ij}|$
($ {\bf r}_{ij}= |{\bf r}_i -  {\bf r}_j|$).
Then for a given
set of vectors $\left( {\1 {\hat r}}_{ij} , \1 k_i, \1 k_j \right)$
for the Lorentz contracted rigid spheres of radius $R_{\rm o}$
there exists the minimum distance between their centers
$r_{ij} ({\bf {\hat r}}_{ij}; {\bf k}_i, {\bf k}_j) = {\rm min}|\1 r_{ij}|$.
The dependence of the  potentials
$u_{ij}$ on the coordinates ${\bf r}_i,{\bf r}_j$
and momenta ${\bf k}_i, {\bf k}_j$
can be given in terms of the minimal distance
as follows
\begin{equation} \label{ureli}
u({\bf r}_{i},{\bf k}_i; {\bf r}_j,{\bf k}_j)  %
\left\{ \begin{array}{rr}
0\,,  &\hspace*{0.3cm}|\1 r_i - \1 r_j| >   r_{ij} \lp
{\1 {\hat r}}_{ij}; \1 k_i, \1 k_j
 \rp  \,, \\
 & \\
\infty\,,  &\hspace*{0.3cm}|\1 r_i - \1 r_j| \le  r_{ij} \lp
{\1 {\hat r}}_{ij}; \1 k_i, \1 k_j
 \rp  \,.
\end{array} \right.
\end{equation}

The general approach to the cluster and virial expansions \cite{MGM:77} 
is valid for this momentum dependent potential, and
in the grand canonical ensemble 
it leads to the transcendental equation for pressure \cite{Bug:00} 
\begin{equation} \label{vdwgcp}
p(T,\mu)~=~T\rho_t(T)~\exp{\lp \frac{\m - a_2p}{T} \rp }~\equiv
~p_{id}(T,\mu-a_2p)~,
\end{equation} 

\noindent
with the second virial coefficient
\begin{eqnarray} 
a_{2} (T) & = &
\frac{g^2}{ \rho_t^2}
\int
\frac{ d{\1 k}_{1}
d{\1 k}_{2} }{(2\pi)^6}
\,e^{\textstyle - \frac{ E( k_{1}) + E( k_{2}) }{T} }
\,\, v (\1 k_1, \1 k_2)  \,\,, 
\label{aiireli}
\\
\label{vreli}
v(\1 {k}_1, \1 {k}_{2}) & = &
\frac{1}{2}
~\int d{\bf r}_{12}~\Theta\left(r_{12}
({\bf {\hat r}}_{12}; {\bf k}_1, {\bf k}_2)~-
~|{\bf r}_{12}|\right)~,
\end{eqnarray}

\noindent
where the thermal density is defined as follows 
$\rho_t(T) = g \int \frac{ d{\1 k} }{(2\pi)^3} e^{\textstyle - \frac{ E( k)}{T} }$, 
\mbox{degeneracy as $g$,} and $v(\1 {k}_1, \1 {k}_{2})$ denotes the relativistic analog of the 
usual excluded volume for the two spheres 
moving with the momenta $\1 k_1$ and $\1 k_2$ 
and, hence, the factor $1/2$ in front of the volume integral in \req{vreli} accounts for the fact that 
the excluded volume of two moving spheres is taken per particle.

In what follows we do not include the antiparticles into consideration to keep it simple,
but this can be done easily. 
Then
the pressure  \req{vdwgcp} generates the following particle density
\begin{equation} \label{iifive}
n (T, \mu)  =  
 \frac{ \partial p (T, \mu) }{ \partial \mu }   =   
 \frac{ e^{\textstyle \frac{\m}{T} }  \rho_t(T)  }{ 1 + e^{\textstyle \frac{\m}{T} }  \rho_t(T)  a_{2} (T)   } \equiv
 \frac{ p}{T \lp 1 + e^{\textstyle \frac{\m}{T} }  \rho_t(T)  a_{2} (T)    \rp } \,,
\end{equation} 
which in the limit of high pressure $p (T, \mu) \rightarrow \infty$ gives a limiting value of particle density 
$n (T, \mu)  \rightarrow a_{2}^{-1} (T)$. 

A form of Eq.~\req{vdwgcp} with constant $a_2$ was  obtained for the first time  in Ref. \cite{Ris:91}.
The new feature of Eq.~\req{vdwgcp} is the temperature dependence of the excluded
volume $a_2$ (T)  \req{aiireli} which is due to the Lorentz contraction
of the rigid spheres. 
This is a necessary and important modification which accounts for
the relativistic properties of the interaction. 
It leads, for instance, to a 50 \% reduction of the excluded volume of 
pions already at temperatures $T = 140 $ MeV \cite{Bug:00}.

The calculation of the cluster integral in relativistic case is more complicated
because each sphere becomes an ellipsoid due to the Lorentz contraction and because the 
relativistic excluded volume strongly depends not only  on the contraction of the spheres, but also 
on the angle between the particle 3-velocities.
Therefore, in Appendix A we give a derivation of  a rather simple formula 
for the coordinate space integration in $a_2$ which is   
found to be 
valid with an accuracy of a few percents for all temperatures. 
Its simplicity 
enables us to perform the angular integrations in $a_2 (T)$ analytically and obtain 
\begin{equation} \label{aiirelii}
a_2 (T) \approx \frac{\a v_{\rm o} }{8} \lp  3 \pi + 
\frac{74\, \rho_s}{3\, \rho_t} \rp\,\,, \quad
\rho_s (T) = \int  \frac{ d {\1 k} }{(2\pi)^3} \frac{m}{ E }
\,\,e^{\textstyle - \frac{ E }{T} } \,\,.
\end{equation} 

\noi
The expression for the 
coefficient $\a \approx 1/ 1.065 $  is given in Appendix A by Eq.  \req{vcorr}.
Using this result it is easy to show that
in the limit of high temperature $T \gg m$ the ratio of  the  scalar density $\rho_s(T)$ to  the thermal 
density $\rho_t (T)$ in \req{aiirelii} vanishes and the second virial coefficient approaches the constant value:
\begin{equation} 
a_2 (T) \biggl|_{T \gg m} \longrightarrow \,\,\frac{ 3 \pi \a v_{\rm o} }{8}  +
{\rm O}\lp \frac{m}{T} \rp \,\,, \biggr. \quad
\end{equation} 
\label{aiireliii}

\noi
which is  about $\frac{3\pi}{32}$ times smaller  compared  to
the value of the nonrelativistic excluded volume, 
and, hence, is surprisingly very close to the dense packing
limit of the nonrelativistic hard spheres.
Similarly to the nonrelativistic VdW case \cite{Raju:92} 
this leads to the problem with causality at very high pressures.
Of course, in this formulation the superluminar speed of sound
should  appear at  very high  temperatures which   are unreachable in hadronic phase. 
Thus the simple  ``relativization''  of the  virial expansion is much more realistic 
than the nonrelativistic description used in Refs. \cite{Yen:99,St:99}, but it   does not
solve the problem completely.

The reason why the simplest generalization \req{vdwgcp}  fails is rather trivial.
Eq. \req{vdwgcp} does not take into account the fact that at high densities
the particles disturb the motion of their neighbors.
The latter leads to the more compact configurations than predicted by
Eqs. \mbox{(\ref{vdwgcp} - \ref{vreli}),} i.e.,
the motion of neighboring  particles becomes
correlated due to a simple geometrical reason.
In other words, since the $N$-particle distribution
is a monotonically decreasing function of the
excluded volume, the most probable state
should correspond to the configurations of smallest
excluded volume of all neighboring particles.
This subject is, of course, far beyond the present paper. 
Although we will touch this subject slightly while discussing  the limit $\mu/T \gg 1$ in Sect. 3,
our primary task here will be  to give  a 
relativistic generalization of the VdW EOS,
which at low pressures behaves in accordance with the relativistic 
virial expansion presented above, and  at the same time  is  free
of the causality paradox at high pressures.  

In our treatment, we will  completely neglect the angular rotations of the Lorentz contracted spheres
because their correct analysis can be done only  within the framework of
quantum scattering theory which is beyond the current scope.
However, it is clear that the rotational effects can be safely neglected at low densities because there are
not so many collisions in the system. At the  same time the rotations of the Lorentz contracted spheres
at very high pressures, which are of the principal interest, can be neglected too, because at
so high densities the particles should be 
so close to each other, that they  must  prevent the rotations of   neighboring particles. 
Thus,  for these two limits we can safely ignore the  rotational effects and proceed further on like 
for  the usual VdW EOS. 

Eq. \req{vdwgcp} is only one of many possible VdW extrapolations to high density.
As in nonrelativistic case,
one can write many expressions which will give the first two terms
of the full virial expansion exactly, and
the difference will appear in the third virial coefficient.
In relativistic case there is an additional ambiguity:
it is possible to perform the momentum integration, first, and 
make the VdW extrapolation next, or vice versa. The result will,
evidently, depend on the order of operation.

As an example let us give a brief ``derivation'' of Eq. \req{vdwgcp}, and its 
counterpart in the grand canonical ensemble. The two first terms of the standard  cluster expansion 
read as
\cite{MGM:77,Bug:00}  
\begin{equation} 
p =  T \, \rho_t (T) \,\,e^{\textstyle \frac{\m }{T} }   
\lp  1 - a_2 \, \rho_t (T) \, e^{\textstyle \frac{\m }{T} }  \rp 
\,\,. 
\end{equation} 
\label{presi}

\noi
Now we approximate the last term on the  right hand side as  
$\rho_t (T) \, e^{\textstyle \frac{\m }{T} } \approx  \frac{p}{T}$. Then we 
extrapolate it to high pressures by  moving this term into the  exponential  function as
\begin{equation} \label{presii}
p \approx  T \, \rho_t (T) \,\,e^{\textstyle \frac{\m }{T} }
\lp  1 - a_2 \,  \frac{p}{T}  \rp 
\approx T \, \rho_t (T) \, \exp\lp \frac{\m - a_2\,p }{T}  \rp 
\,\,.
\end{equation} 

\noi
The resulting expression coincides with Eq. \req{vdwgcp}, but the above manipulations make it simple
and transparent. 
Now we will repeat all  the above steps while keeping
both momentum integrations fixed 
\begin{eqnarray} 
p & \approx & \frac{T\, g^2\, e^{\textstyle \frac{\m }{T}  }}{ \rho_t (T)}
\int
\frac{d{\1 k_1}}{(2\pi)^3}
\frac{d{\1 k_2}}{(2\pi)^3}
\,\,e^{\textstyle - \frac{E(k_1) + E(k_2) }{T} }
\lp  1 - \frac{ v (\1 k_1, \1 k_2)\, p }{T}   \rp
\nonumber \\
\label{presiii}
& \approx & \frac{T\, g^2 }{ \rho_t (T)}
\int
\frac{d{\1 k_1}}{(2\pi)^3}
\frac{d{\1 k_2}}{(2\pi)^3}
\,e^{\textstyle \frac{\m - v (\1 k_1,\1 k_2)\, p\, -\,E(k_1)\, -\,E(k_2) }{T} }
\,\,.
\end{eqnarray} 

The last expression contains the relativistic excluded volume \req{vreli} explicitly
and, as can be shown, is free of the causality paradox. 
This is so because at high pressures the main contribution to the momentum 
integrals corresponds to the  smallest values of the excluded volume \req{vreli}.
It is clear  that such  values are reached when the both spheres are ultrarelativistic and 
their velocities are collinear.

With the help of the following notations for the averages
\begin{eqnarray}
\bra {\cal O} \ket  & \equiv  & \frac{ g }{ \rho_t (T) } \int \frac{d{\1 k}}{(2\pi)^3} \,\, {\cal O}  \,\,e^{\textstyle - \frac{E(k) }{T} } \,,
\\
\bra\bra {\cal O} \ket \ket & \equiv  & \frac{ g^2 }{ \r^2_t (T) } \int \frac{d{\1 k_1}}{(2\pi)^3} \frac{d{\1 k_2}}{(2\pi)^3}
\,\, {\cal O}  \,\,e^{\textstyle - \frac{ v (\1 k_1,\1 k_2)\, p\, +\,E(k_1)\, +\,E(k_2) }{T} } \,,
\label{presiiii}
\end{eqnarray} 
we can define all other thermodynamic functions as 
\begin{eqnarray} \label{iiizeroA}
n (T, \m) & = & 
 \frac{ \partial p (T, \m) }{ \partial \m }   =   \frac{ p}{T \lp 1 + e^{\textstyle \frac{\m}{T}}  \rho_t(T) 
  \bra\bra  v(\1 k_1, \1 k_2)  \ket \ket \rp } \,, \\
s (T, \m) & = & 
 \frac{ \partial p (T, \m) }{ \partial T }   =  \frac{ p}{T} +  \frac{ 1}{T} 
 \frac{ \lp 2\,e^{\textstyle \frac{\m}{T}}  \rho_t(T)  \bra\bra E \ket \ket  - [\mu +  \bra E \ket ]\, p\, T^{-1}  \rp}{
 1 + e^{\textstyle \frac{\m}{T}}  \rho_t(T)  \bra\bra v(\1 k_1, \1 k_2) \ket \ket } \,, \\
\varepsilon (T, \m) & = & 
T\,s (T, \m) + \m\,n (T, \m) - p (T, \m)  =  \frac{  2\,e^{\textstyle \frac{\m}{T}}  \rho_t(T) 
 \bra\bra E \ket \ket  - [ \mu +  \bra E \ket ] \, p\, T^{-1}  }{
 1 + e^{\textstyle \frac{\m}{T}}  \rho_t(T)  \bra\bra  v(\1 k_1, \1 k_2)  \ket \ket } \,.
\end{eqnarray}
Here  $n (T, \m)$ is the particle density, while  $s (T, \m)$ and $\varepsilon (T, \m) $ denote the entropy 
and energy density, respectively.

In the low pressure limit $ 4 \, p \, v_{\rm o}  T^{-1} \ll 1$ the corresponding exponent  in 
\req{presiiii} can be expanded and the mean value of the relativistic excluded volume can be 
related to the second virial coefficient  $a_2 (T)$ as follows
\begin{equation} \label{iiizeroB}
\bra\bra v(\1 k_1, \1 k_2) \ket \ket  \approx   a_2 (T) ~ - ~ 
\frac{p}{T} \bra\bra v^2(\1 k_1, \1 k_2) \ket \ket  \, ,
\end{equation} 
which shows that at low pressures the average value of the relativistic excluded volume
should match the second virial coefficient  $a_2 (T)$, but should be smaller than $a_2 (T)$ at 
higher pressures and this behavior is clearly seen in Fig. 1.

%%%%%%%%%%%%%%%%%%%%%%%%

\begin{figure}

\mbox{
\hspace*{3.0cm}\epsfig{figure=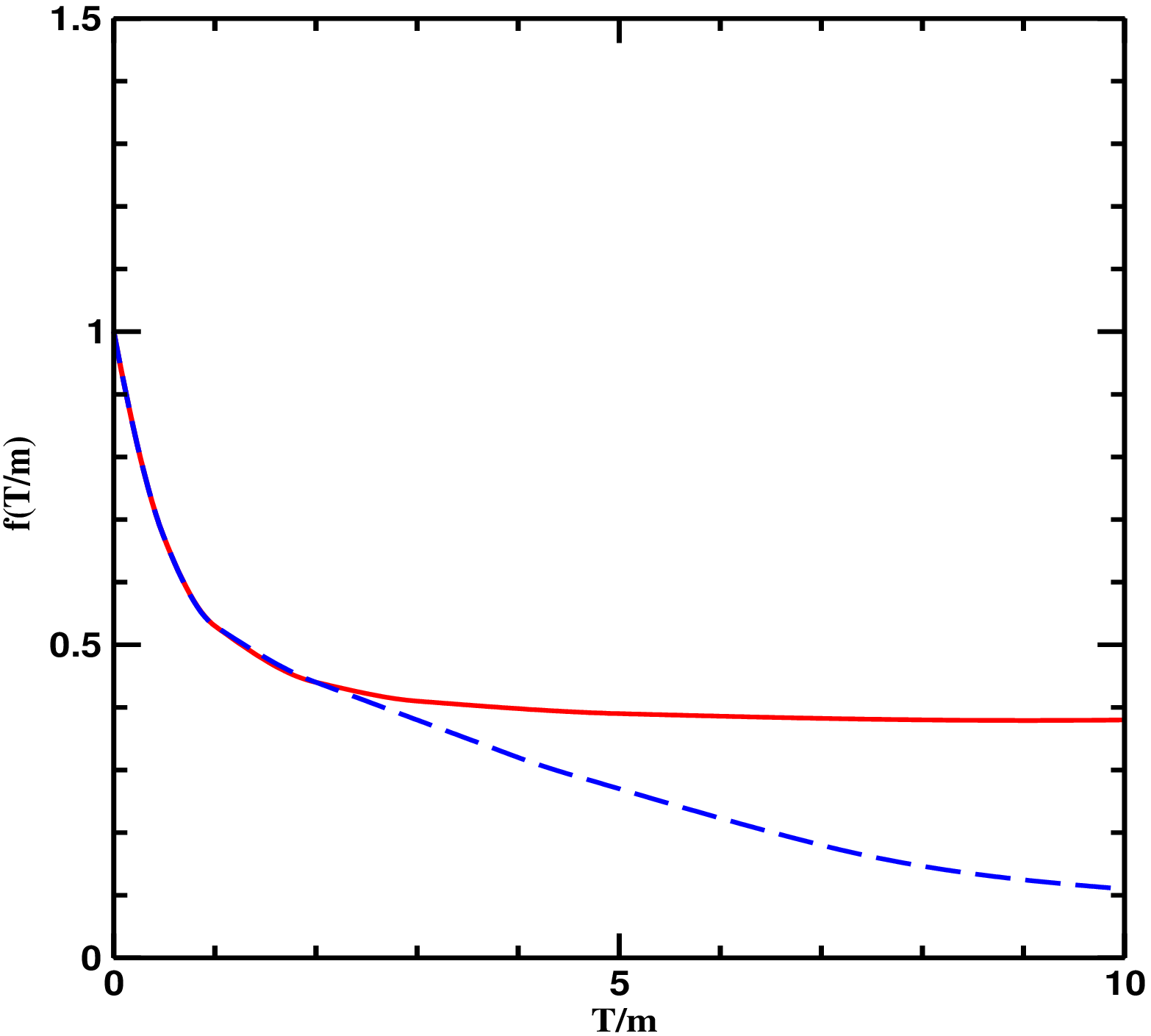,height=8.8cm,width=9.45cm}}

\vspace*{0.5cm}

{\bf Fig. 1.}  Comparison of the exact value of the second virial coefficient $a_2 (T)/a_2(0)$ (solid curve)
 with the averaged    value of the relativistic excluded volume 
$\alpha \bra\bra  v (\1 k_1, \1 k_2)  \ket \ket / ( a_2(0))$ (dashed curve) given by Eq. \req{iiione}
fot $\mu = 0$. 
The  normalization coefficient 
$\alpha \approx 1/1.065$  \req{vcorr} is introduced  to reproduce the low density results.

\end{figure}

%%%%%%%%%%%%%%%%%%%%%%%%%

\vspace*{0.0cm}

A comparison of the particle densities \req{iifive} and \req{iiizeroA} shows that despite the different 
formulae  for pressure the particle densities of these models have a very similar expression, but  in \req{iiizeroA} 
the second virial coefficient
is replaced by  the averaged value of the relativistic excluded volume $  \bra\bra  v(\1 k_1, \1 k_2)  \ket \ket  $.
Such a  complicated dependence of the particle density  \req{iiizeroA} on  $T$ and $\mu$ requires a nontrivial  analysis for  the limit of  high pressures.  

To analyze  the high pressure limit $p \rightarrow \infty$ analytically we need an analytic  expression for  
the excluded volume. For this purpose we will use the ultrarelativistic expression derived in the Appendix A: 
\begin{equation} \label{iiione}
v(\1 k_1, \1 k_{2})  \approx \frac{v^{Urel}_{12}(R, R)}{2} \equiv   \frac{  v_{\rm o} }{2 }  
\lp  \frac{m}{E(\1 k_1)} + \frac{m}{E(\1 k_2)} \rp
\lp  1 + \cos^2 \lp \frac{\Theta_v}{2} \rp \rp^2  
 +  \frac{3 \, v_{\rm o} }{2}  \sin \lp \Theta_v \rp   \,\,.
\end{equation} 
As usual, the total excluded volume $v^{Urel}_{12}(R, R)$ is taken per particle. 
Eq. \req{iiione} is valid for $0 \le \Theta_v \le \frac{\pi}{2} $; to use it for 
$ \frac{\pi}{2} \le \Theta_v \le \pi $ one has to make a replacement 
$ \Theta_v \longrightarrow \pi -  \Theta_v$ in  \req{iiione}. 
Here  the coordinate system  is chosen in such a way that the angle $\Theta_v$ between the 3-vectors of 
particles' momenta $\1 k_1$ and  $\1 k_{2}$ coincides with the usual spherical angle $\Theta$ of  spherical
coordinates (see Appendix A). 
To be specific,  the OZ-axis of the momentum space  coordinates of  the second particle  is chosen to coincide with the 3-vector of the momentum $\1 k_1$ of the first particle.

The Lorentz frame is chosen to be the rest frame of the whole system because 
otherwise the expression for pressure becomes cumbersome.  Here $v_{\rm o}$ stands for the eigen volume of particles which, for simplicity, are assumed to
have the same hard core radius and the same mass. 

Despite the fact that this 
equation  was obtained for  ultrarelativistic limit, it is to a within few per cent accurate in the whole range of parameters
(see Fig. 1 and Appendix A for the details),
and, in addition,  it is  sufficiently simple to allow the analytical treatment.

\vspace*{0.5cm}

%%%%%%%%%%%%%%%%%%%%%%%%%

\begin{figure}

\mbox{
\hspace*{0.0cm}\epsfig{figure=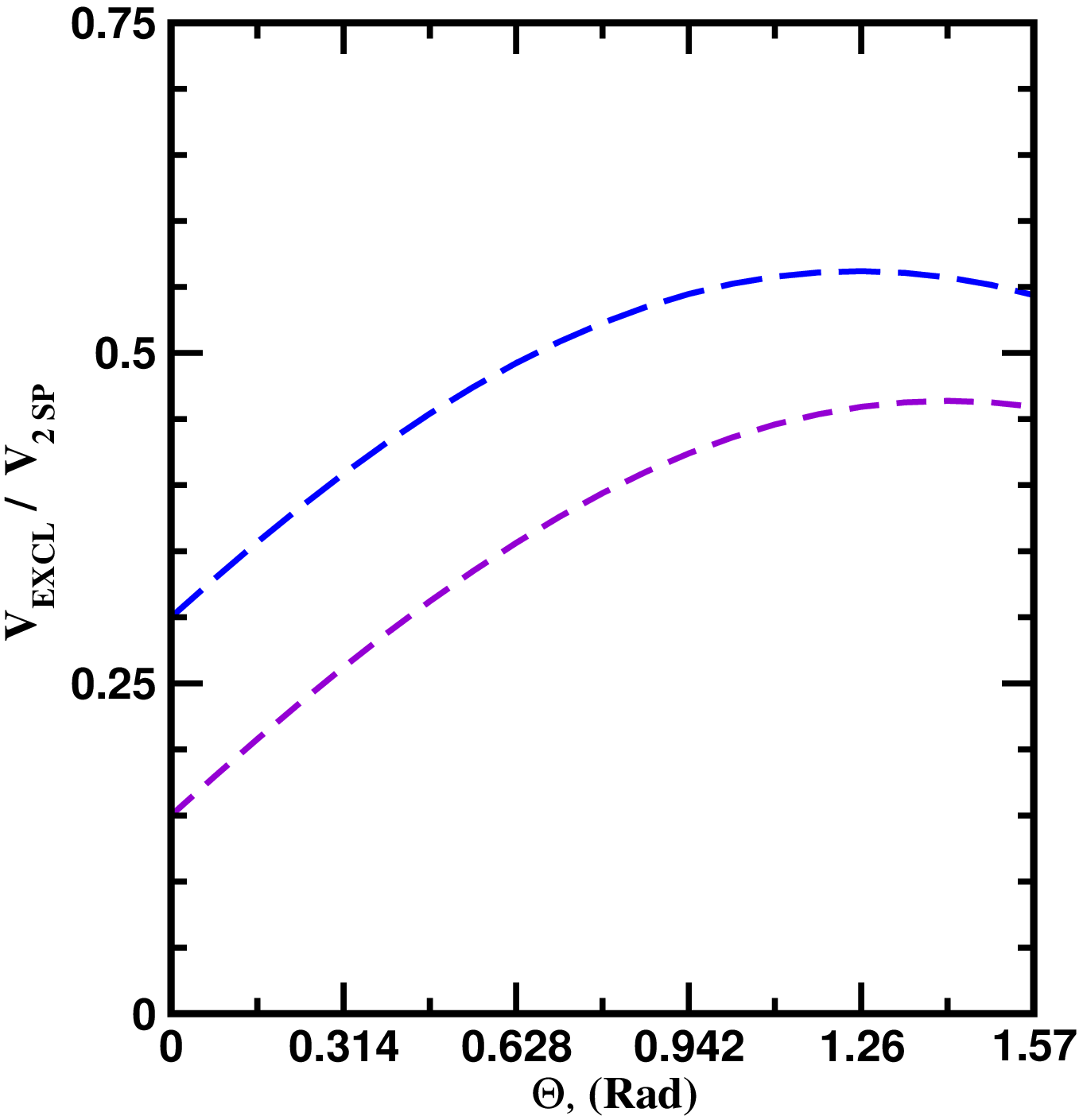,height=7.0cm,width=8.cm} 
\hspace*{-0.5cm} \epsfig{figure=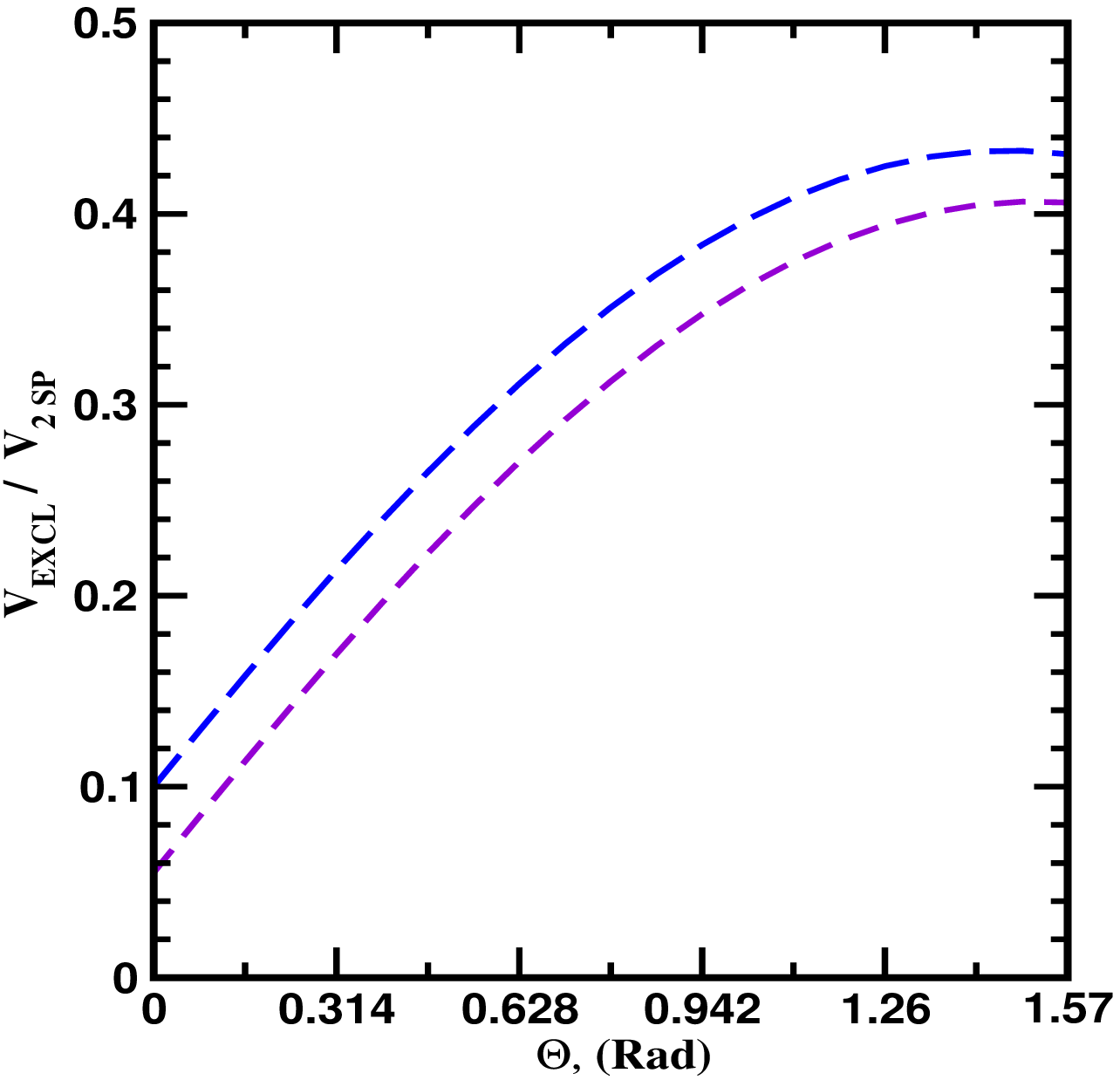,height=7.0cm,width=8.cm}
}

\vspace*{0.3cm}

{\bf Fig. 2.}
Comparison of the relativistic  excluded volumes 
for highly contracted spheres. 
In the left panel the long dashed curve corresponds  to 
$\frac{E(k_1)}{m} = 2$ and  $\frac{E(k_2)}{m} = 10$ whereas 
the short dashed curve is found for  
$\frac{E(k_1)}{m} = 5$ and  $\frac{E(k_2)}{m} = 10$.
The corresponding values in the right panel 
are $\frac{E(k_1)}{m} = 10$,  $\frac{E(k_2)}{m} = 10$ (long dashed curve) and  
$\frac{E(k_1)}{m} = 10$,   $\frac{E(k_2)}{m} = 100$ (short dashed curve).
It shows that the  excluded volume for  $\Theta_v$ close to $\frac{\pi}{2}$ is 
finite always, while for the collinear velocities the excluded volume   approaches zero,  
if  both spheres are ultrarelativistic. 
\end{figure}

\vspace*{0.4cm}

\bc
{\large \bf 3.  High Pressure Limit}
\ec

\vspace*{0.2cm}

As seen from the  expression for the relativistic excluded volume \req{iiione} , 
for very high pressures only the  smallest values of the relativistic excluded volume will 
give a non-vanishing contribution to  the angular integrals of thermodynamic functions.
This means that only  $\Theta_v$-values  around $0$ and around $\pi$ will contribute  into the  thermodynamic functions (see Fig. 2). Using the variable $x = \sin^2 \lp \Theta_v/2 \rp$, one can rewrite the
$\1 k_{2}$ angular integration as follows
\begin{eqnarray} \label{iiitwo}
\hspace*{-0.5cm}&&I_{\Th} ( k_1)  =  
\int
\hspace*{-0.1cm}
\frac{d{\1 k_2}}{(2\pi)^3} 
e^{\textstyle - \frac{  v ( \1 k_1, \1 k_2) p  }{T} } 
4 \int
\hspace*{-0.1cm}
\frac{d\,k_2  k^2_2 }{(2\pi)^2} 
\int_0^{0.5} \hspace*{-0.1cm} d\,x
\,\,e^{ -   \lp A C \lp 1 - \frac{x}{2} \rp^2  + B \sqrt{x (1 - x )} \rp   }
\,\,,\, \quad\quad \\
\label{iiithree}
\hspace*{-0.5cm}&&{\rm with}\hspace*{0.5cm}    A = 2 v_{\rm o}  \frac{p}{T} \,;
\hspace*{0.3cm} B =  \frac{3}{2} A\,;
\hspace*{0.3cm} C = 
\lp \frac{m}{E( k_1)} +  \frac{m}{E( k_2)} \rp\,\,,
\end{eqnarray}
where we have  accounted for the fact  that  the integration over the polar angle gives a factor $2\pi$
and that one should double the integral value in order to integrate  over a half of the $\Theta_v$ range.

Since $C \le 2$ in \req{iiithree} is a decreasing function of the momenta, then in the limit $A \gg 1$ one can 
account only for  the $\sqrt{x}$ dependence in the exponential in \req{iiitwo} because it is the leading one. 
Then integrating by parts one obtains 
\begin{equation} \label{iiifour}
\hspace*{-0.5cm}I_{\Th} ( k_1) \approx 
4 \int
\hspace*{-0.1cm}
\frac{d\,k_2  k^2_2 }{(2\pi)^2}
\,\,e^{ -  A C}
\int_0^{0.5} \hspace*{-0.1cm} d\,x
\,\,e^{ -  B \sqrt{x }    }
\approx
8 \int
\hspace*{-0.1cm}
\frac{d\,k_2  k^2_2 }{(2\pi)^2}
\,\,e^{ -  A C} \frac{1}{B^2}
\,\,.\, 
\end{equation}

Applying the above result  to the pressure \req{presiii}, in the limit under consideration  one finds 
that the momentum integrals are decoupled and  one gets the following equation for pressure
\begin{equation} \label{iiifive}
\hspace*{-0.5cm} p (T, \m)  \approx  \frac{16\, T^3 e^{\frac{\mu}{T}}  }{9 \,v_{\rm o}^2 \, p^2 \, \rho_t(T) }
\left[  g  \int
\hspace*{-0.1cm}
\frac{d\,k  k^2}{ (2 \pi)^2 }
\,\,e^{\textstyle - \frac{E(k)}{T} - \frac{2 \, v_{\rm o} \, m  }{ T E(k)}\, p  }
\right]^2   \,.
\end{equation}
Now it is clearly  seen that at high pressures the momentum distribution function in  (\ref{iiifive}) may essentially differ from the Boltzmann one.  To demonstrate this we can calculate an effective  temperature by differentiating the exponential under the integral in (\ref{iiifive})  with respect to particle's energy $E$:
\begin{equation} \label{iiifivet}
\hspace*{-0.5cm} T_{eff} (E)  ~=~  - \left[ \frac{\partial}{\partial E} \left( - \frac{E}{T} - \frac{2 \, v_{\rm o} \, m  }{ T E }\, p     \right) \right]^{-1} ~=~  \frac{T}{ 1 ~ - ~ \frac{2 \, v_{\rm o} \, m  }{E^2} p   }~.
\end{equation}
Eq. (\ref{iiifivet}) shows that the effective temperature $T_{eff} (E \rightarrow \infty) = T$ 
may be essentially lower  than  that one at  $E = m$. In fact, at very high pressures the 
effective temperature $T_{eff} (m)$  may become negative.   

A sizable  difference between  $T_{eff}$ values at  high and low particle energies mimics the collective motion of particles since a similar behavior is typical for the transverse energy spectra 
of  particles having the collective transverse velocity which monotonically grows  with the transverse  radius \cite{Bugaev:02a,Bugaev:02b}. However, in contrast  to  the true 
collective motion case  \cite{Bugaev:02a,Bugaev:02b} ,
the low energy $T_{eff}$  (\ref{iiifivet})  gets higher for smaller 
masses of particles.  Perhaps, such a different behavior of low energy  effective temperatures 
can be helpful  for distinguishing  the   high pressure case from the collective motion of particles.

Our next step is to perform the gaussian integration in  Eq.  \req{iiifive}.
Analyzing the function 
\begin{equation} \label{iiisix}
F \equiv 2 \ln k - \frac{E (k)}{T} - A  \frac{m}{E (k)} \,
\end{equation} 
for  $A \gg 1$, one can safely use the ultrarelativistic approximation for particle momenta  
$k \approx E(k) \rightarrow \infty $. Then it is easy to see  that 
the function F in  \req{iiisix} has an extremum at 
\begin{equation} \label{iiiseven}
\frac{\partial F}{\partial E} =   \frac{2}{E} - \frac{1}{T} + A  \frac{m}{E^2} = 0 \quad  \Rightarrow \quad E = E^*  \approx  \frac{A\, m}{\sqrt{1 + \frac{A\, m}{T} }  - 1 }  \equiv  T \lp  \sqrt{1 + \frac{A\, m}{T} } + 1 \rp \,,
\end{equation}
which turns out to be a maximum, since the second derivative of F   \req{iiisix} is negative
\begin{equation} \label{iiieight}
\biggl. \frac{\partial^2 F}{\partial E^2}\biggr|_{E = E^*} \approx  -  \frac{2}{(E^*)^2} - 2\,A  
\frac{m}{(E^*)^3}  < 0
 \,.
\end{equation}

There are two independent ways to increase  pressure: one can increase 
the value of chemical potential while keeping
temperature fixed and vice versa.  We will consider the high chemical potential limit $\mu/T \gg 1$ for finite 
$T$ first,  since this case is rather unusual.  In this limit the above expressions can be simplified further on 
\begin{equation} \label{iiieightB}
E^* \approx \sqrt{2\, m\, v_{\rm o} \, p }\,, \quad \Rightarrow \quad  
\biggl. \frac{\partial^2 F}{\partial E^2}\biggr|_{E = E^*} \approx -  \frac{2}{T \, \sqrt{2\, m\, v_{\rm o} \, p } } \,.
\end{equation}
Here in the last step we explicitly substituted   the  expression for $A$. 
Performing the gaussian integration for momenta in \req{iiifive}, one arrives at
\begin{equation} \label{iiinine}
\int
\hspace*{-0.1cm}
\frac{d\,k  k^2}{ (2 \pi)^2 }
\,\,e^{\textstyle - \frac{E(k)}{T} - \frac{2 \, v_{\rm o} \, m  }{ T E(k)}\, p  } \approx
\frac{(E^*)^2}{ (2 \pi)^2 }\, \sqrt{\pi \, T \, E^*}\, e^{\textstyle - \frac{2 \, E^*}{T} }\,,
\end{equation}
which leads to the following equation for the most probable energy of particle $E^*$
\begin{equation} \label{iiiten}
E^* \approx  D\, T^4 \,  \,e^{\textstyle  \frac{\mu - 4 E^*}{T} }\,, \quad 
D \equiv \frac{ 8\, g^2 m^3 v_{\rm o}}{9\, \pi^3 \rho_t(T)}\,.
\end{equation}
As one can see, Eq. \req{iiiten} defines pressure of the system. Close inspection shows that the high pressure limit 
can be achieved, if the exponential in \req{iiiten}  diverges much slower than $\mu/T$. The latter defines the 
EOS in the leading order as 
\begin{equation} \label{iiieleven}
E^* \approx  \frac{\mu}{4}\,, \quad \Rightarrow \quad  
p \approx \frac{\mu^2}{32\, m \, v_{\rm o}}  \,.
\end{equation}
The left  hand side equation above demonstrates that in the $\mu/T \gg 1$ limit the natural energy scale is given by a chemical potential.  This is a new and important feature  of the relativistic VdW EOS compared to the previous findings.  

The right hand side Eq. \req{iiieleven}
allows one to find all other thermodynamic functions in this limit  from thermodynamic  identities:
\begin{eqnarray} \label{iiitwelve}
s  \approx  0 \,, \quad
n  \approx  \frac{2 \, p}{\mu} \,, \quad
\varepsilon  \equiv  T s + \mu n - p \approx p \,.
\end{eqnarray} 
Thus, we showed that for $ \mu/T \gg 1$  and finite $T$  the speed of sound $c_s$ in the leading order does 
not exceed the speed of light since
\begin{equation} \label{iiithirteen}
c_s^2 =  \biggl. \frac{\partial p}{\partial \varepsilon}\biggr|_{s/n} =  \frac{d\, p}{d\, \varepsilon} = 1\,.
\end{equation}
From Eq. \req{iiiten} it can be shown  that the last result holds  in all orders. 

It is interesting  that the left hand side  Eq.  \req{iiieightB} has a simple kinetic interpretation.  
Indeed, recalling that the pressure is the change of momentum during the collission time one can write  \req{iiiseven} as follows
(with $E^* = k^*$)
\begin{equation} \label{iiifourteen}
p =  \frac{(k^*)^2}{ 2\, m\,  v_{\rm o}} =  \frac{2\, k^*}{\pi R_{\rm o}^2}\cdot \frac{3\,v^* \gamma^*}{8\, R_{\rm o}} \cdot  \frac{1}{2}\,.
\end{equation}
In the last result the change of momentum during the collision with the wall  is $2\, k^*$, which takes the time 
$\frac{8\, R_{\rm o}}{3\,v^* \gamma^*}$. The latter is  twice of the Lorentz contracted height ($4/ 3 R_{\rm o}$)  of  the cylinder of the base $\pi R_{\rm o}^2$ which is  passed  with the speed $ v^* $.  
Here the particle velocity $v^*$ and the corresponding gamma-factor 
$\gamma^*$ are defined as $ v^* \gamma^* = k^* / m$. The rightmost factor $1/2$ in \req{iiifourteen}
accounts  for the fact that only a half of particles moving perpendicular to the wall has the momentum $- k^*$.   Thus, Eq. \req{iiifourteen} shows that in the limit under consideration the pressure is generated by the particle momenta which are perpendicular
to the wall.
This, of course, does not mean that all particles in the system have the momenta which are perpendicular to a
single wall. 
No, this means  that  in those places near the wall where the particles' momenta are not perpendicular (but are parallel) to it, 
 the  change of   momentum  $2 k^*$ is transferred  to the wall  by  the particles  located in  the inner regions of the system  whose 
momenta are perpendicular to the wall. 
Also it is easy to deduce that such a situation is possible, if 
the system is divided into the rectangular cells or boxes inside which the particles are moving along the height of the box and  their  momenta  are collinear, but they are  perpendicular 
to the
particles' momenta in all surrounding  cells.  Note that appearing of particles' cells is a typical feature  of the treatment of high density limit \cite{Muen} and can be related to  a complicated 
phase structure of nuclear matter at  very low temperatures \cite{Gyulassy:85}. 

Of course,  inside of  such a box each Lorentz contracted  sphere would generate an excluded volume which is  equal to a volume of a cylinder of height  $\frac{2\, R_{\rm o}}{\gamma^*} $ and base  $\pi R_{\rm o}^2$. This cylinder, of course,  differs from the cylinder involved in Eq. \req{iiifourteen}, but we note that exactly the hight $\frac{4 R_{\rm o}}{3\, \gamma^*}$ is used in the derivation of the ultrarelativistic  
limit for  the relativistic excluded volume \req{vunnor} (see Appendix A for details).
Thus, it is very interesting that in contrast to 
nonrelativistic case the relativistic excluded volume $\frac{4 \pi  R_{\rm o}^3}{3\, \gamma^*}$ which enters into  Eq. \req{iiifourteen}  is only 33~\% smaller than the excluded volume   
$\frac{2 \pi  R_{\rm o}^3}{ \gamma^*}$  of ultrarelativistic particle at high pressures.
Also  
 it is remarkable that the  low density  EOS extrapolated to very high values of the chemical potential, at which it is not supposed to be valid at all,  gives  a reasonable estimate  for the pressure at high densities. 

Another interesting conclusion that follows  from this limit is   that  for the relativistic VdW  systems existing in the nonrectangular 
volumes the relativistic analog of the dense packing  may be  unstable.

The analysis of the limit $T/\mu \gg 1$ and finite $\mu$ also starts from Eqs. \req{iiifive}--\req{iiiseven}.
The function F from \req{iiisix} again has the maximum at $E^* \equiv E (k^*) = k^*$  defined by the 
right hand side Eq. \req{iiiseven}.
Now  the second derivative of function $F$ becomes 
\begin{equation} \label{iiisixteen}
\biggl. \frac{\partial^2 F}{\partial E^2}\biggr|_{E = E^*} \approx  -  \frac{2}{(E^*)^2} - 2\,A  \frac{m}{(E^*)^3} = 
-  \frac{2 \, \sqrt{ 1 + \frac{A\, m}{T}   } }{ (E^*)^2}  \,.
\end{equation}
This result  allows one to perform the gaussian integration for momenta in \req{iiifive} for this limit and get 
\begin{equation} \label{iiiseventeen}
\int
\hspace*{-0.1cm}
\frac{d\,k  k^2}{ (2 \pi)^2 }
\,\,e^{\textstyle - \frac{E(k)}{T} - \frac{2 \, v_{\rm o} \, m  }{ T E(k)}\, p  } \approx
\frac{(E^*)^3 \, e^{\textstyle -  2 \, \lp 1 + \frac{A\, m}{T}   \rp^{\frac{1}{2}}  } }{ (2 \pi)^2\, 
\lp 1 + \frac{A\, m}{T}   \rp^{\frac{1}{4}} }\,  I_\xi \lp 1 + \frac{A\, m}{T}   \rp \,, 
\end{equation}
where the auxiliary integral $I_\xi$ is defined as follows
\begin{equation} \label{iiieighteen}
 I_\xi (x) \equiv  \int\limits_{- {x}^{\frac{1}{4}}}^{+ \infty} d \xi \,\,e^{- \xi^2} \,.
\end{equation}
The  expression (\ref{iiiseventeen}) can be also used to find  the thermal density $\rho_t(T)$ in the limit $T\rightarrow \infty$ by the substitution $A = 0$. Using \req{iiiseventeen}, one can rewrite the equation for pressure \req{iiifive} as the 
equation for the unknown variable  $z \equiv A\, m / T \equiv  \frac{2 \, v_{\rm o} \, m \, p}{ T^2}$
\begin{equation} \label{iiinineteen}
z^3 \approx \,  e^{\textstyle  \frac{\mu}{T}  } \phi (z)\,, \quad 
\phi (z) \equiv \frac{2\, g\,v_{\rm o} \, m^3\, I_\xi^2 (1+z) \lp 1 + (1 + z )^{\frac{1}{2}} \rp^6  }{
 \lp 3\,\pi \,e^{2 \cdot \sqrt{1+z} - 1}  \rp^2  I_\xi (1)  (1 + z )^{\frac{1}{2}} } \,.
\end{equation}

Before  continuing our analysis  further on, it is necessary to make  two comments concerning 
Eq. \req{iiinineteen}. 
First, rewriting  the left hand side Eq. \req{iiinineteen} in terms of pressure, one can see that
the value of  chemical potential is  formally  reduced exactly in three times. In other words, it looks like that in the limit of high temperature and finite $\mu$ the pressure of the relativistic VdW gas is created by the particles
with the charge being equal to the one third of their original charge. 
Second,  due to the nonmonotonic  dependence of  $\phi(z)$ in the right hand side Eq. \req{iiinineteen} it is
possible that the left hand side Eq. \req{iiinineteen} can have several solutions for some values of parameters.   
Leaving aside the discussion of this possibility,  we will further  consider  only 
such a solution of \req{iiinineteen}  which corresponds to the 
largest value of  the pressure \req{iiifive}. 

Since the function $\phi(z)$ does not have any explicit dependence  on $T$ or  $\mu$, one can establish  a very convenient  relation 
\begin{equation} \label{iiitwenty}
 \frac{ \partial z }{ \partial T }   =  -  \frac{ \mu }{  T }  \,  \frac{ \partial z }{ \partial \mu }  
\end{equation}
between the partial derivatives  of  $z$ given by the left hand side Eq. \req{iiinineteen}. Using (\ref{iiitwenty}),
one can calculate all the  thermodynamic functions  from the pressure  $ p = \beta \, T^2 z $ (with $ \beta \equiv (3\, m \, v_{\rm o})^{-1}$)
as follows:
\begin{eqnarray} \label{iiitwone}
n  & \approx &  \beta \, T^2 \, \frac{ \partial z }{ \partial \mu }\,,    \\
 \label{iiitwtwo}
s  & \approx &   \beta \, \left[  2\, T \, z + T^2  \frac{ \partial z }{ \partial T } \right]  = \frac{2\, p - \mu n }{T}  \,,  \\
\varepsilon &  \equiv &   T s + \mu n - p \approx p \,.
 \label{iiitwthree}
\end{eqnarray} 
The last result  leads to  the causality condition  \req{iiithirteen} for the limit $T/\mu \gg 1$ and finite $\mu$.

In fact, the above result can be extended to any $\mu > - \infty$ and any value of $T$ satisfying  the inequality
\begin{equation} \label{iiitwfour}
E^*  \approx   T \lp  \sqrt{1 + z } + 1 \rp \gg m \,,
\end{equation}
which is sufficient to derive Eq. \req{iiinineteen}. To show this, it is sufficient to see that for $z = 0 $ there holds
the inequality $z^3 < e^{\textstyle  \frac{\mu}{T}  } \phi (z)$, which changes to the opposite inequality 
$z^3 > e^{\textstyle  \frac{\mu}{T}  } \phi (z)$ for $z = \infty$. Consequently, for any value  of $\mu $ and $T$
satisfying  \req{iiitwfour} the left hand side Eq. \req{iiinineteen} has at least one solution $z^* > 0$ for which 
one can establish Eqs. \req{iiitwenty}--\req{iiitwthree} and prove the validity of  the  causality condition \req{iiithirteen}.

The  model  \req{presiii} along with the analysis of high pressure limit  can be straightforwardly generalized to include several particle species.  
For the pressure $p (T, \{ \mu_i\} )$ of the mixture of  $N$-species  with  masses  $m_i$ 
$(i = \{1, 2,.., N\})$, degeneracy $g_i$,  hard core radius $R_i$  and chemical potentials $\mu_i$ is defined as a solution of the following equation
\begin{equation} \label{iiizeroC}
p (T, \{ \mu_i\} ) =  \int
\frac{d^3{\1 k_1}}{(2\pi)^3}
\frac{d^3{\1 k_2}}{(2\pi)^3}  
\sum_{i, j = 1}^N  \frac{T\, g_i\, g_j }{ \rho_{tot} (T, \{ \mu_l\}  )} 
\,e^{\textstyle \frac{\mu_i + \mu_j - v_{ij} (\1 k_1,\1 k_2)\, p\, -\,E_i(k_1)\, -\,E_j(k_2) }{T} } \,,
\end{equation} 
where the relativistic excluded volume per particle of species $i$ 
(with the  momentum $\1 k_1$) and $j$ (with the  momentum $\1 k_2$) is denoted as 
$v_{ij} (\1 k_1,\1 k_2)$,  $E_i(k_1) \equiv \sqrt{k_1^2 + m_i^2}$ and 
$E_j(k_2) \equiv \sqrt{k_2^2 + m_j^2}$ are the corresponding energies, and the total 
thermal density is given by the expression 
\begin{equation} \label{iiioneC}
 \rho_{tot} (T, \{ \mu_i\}  ) = \int
\frac{d^3{\1 k}}{(2\pi)^3} 
\sum_{i = 1}^N ~  g_i  
\,e^{\textstyle \frac{\mu_i  -\,E_i(k) }{T} } \,. 
\end{equation} 
The excluded volume  $v_{ij} (\1 k_1,\1 k_2)$  can be accurately  approximated 
by  $\alpha\, v^{Urel}_{12}(R_i, R_j) / 2$ defined by Eqs. (\ref{vunnor}) and 
(\ref{vcorr}).  

The multicomponent generalization (\ref{iiizeroC}) is obtained in the same sequence 
of steps as the one-component expression (\ref{presiii}). The only difference is  in
the definition of the total thermal density (\ref{iiioneC}) which now includes  the 
chemical potentials. 
Note also that the expression (\ref{iiizeroC})  by construction  recovers the  virial expansion up to the  second order 
at low
particle densities,  but 
it cannot be reduced to any of two extrapolations which are   suggested in  \cite{VdWaals} and 
\cite{Gor:99} for  the multicomponent 
mixtures  and    carefully analyzed  in Ref. \cite{Zeeb:02}. Thus, the expression 
(\ref{iiizeroC}) removes the non-uniqueness of the VdW extrapolations to high densities,
if one requires  a  causal behavior in this limit. 

\vspace*{0.5cm}

\bc
{\large \bf 4.  Concluding Remarks }
\ec

\vspace*{0.2cm}

In this work  we proposed a relativistic analog of the VdW EOS  which  reproduces the 
virial expansion for the gas of  the Lorentz contracted rigid spheres at low particle  densities and  
is  causal  at high densities. 
As one can see from the expression for particle density \req{iiizeroA}
and from the corresponding relation for effective temperature \req{iiifivet}
the one-particle momentum distribution function has a more complicated energy dependence than the usual Boltzmann 
distribution function, which would be interesting to check experimentally. 
Such a task involves considerable technical difficulties since the particle spectra measured in high energy nuclear 
collisions involve a strong collective flow which can easily hide or smear the additional 
energy  dependence.  
However, it is possible that  such a  complicated energy dependence of the momentum spectra and 
excluded volumes of  lightest hardons, i.e. pions and kaons, can be  verified  for highly accurate 
measurements, if the collective flow is correctly taken into account. 
The latter adjustment is tremendously complex because it  is related to  the freeze-out 
problem in  relativistic hydrodynamics \cite{Bug:96} or hydro-cascade approach \cite{Bug:02}. 
Another possibility  to study  the effect of Lorentz contraction on the EOS properties is to 
incorporate them into transport models. The first steps in this direction 
have been made  already in \cite{Larionov:07}, but the approximation used in \cite{Larionov:07}
is too crude.

It might be more realistic   to incorporate the developed approach into effective models of 
nuclear/hadronic matter  
\cite{Ris:91,Na:61,Gu:X,chirm} and check the obtained EOS on a huge amount of data 
collected by the nuclear physics of intermediate energies. Since the suggested 
relativization  of  the VdW EOS makes it softer at high densities,  one can hope to  improve 
the description of the  nuclear/hadronic matter  properties (compressibility constant, elliptic flow, effective nucleon masses e.t.c.) at low temperatures and high baryonic 
densities \cite{Danil:05}. 

Also it is possible that the momentum spectra of this type  can help to extend 
the hydrodynamic description  
into the  region of large transversal momenta of hadrons ($p_T > 1.5 - 2 $ GeV)
which are  usually  thought to be too large to 
follow the  hydrodynamic regime  \cite{Heinz:05}. 

Another possibility to validate the suggested model is to study angular  correlations 
of the hard core  particles emitted from the neighboring regions and/or  the enhancement of
the particle yield  of those hadrons occurring  due to coalescence of the constituents with the  short range repulsion.  
As shown above (also see  Fig. 2), the present model  predicts 
that the probability to find the neighboring  particles  with collinear   velocities
is higher than the one with non-collinear   velocities. 
Due to this reason, the coalescence of particles  with the parallel  velocities 
should be enhanced.  This effect amplifies if   pressure  is high  and if particles are
relativistic in the local rest frame.
Therefore, it would be interesting to study the coalescence of any relativistic  constituents 
with hard core repulsion 
(quarks or hadrons)
at high pressures  in a spirit of the recombination model of Ref. \cite{Mull:03} and extend  its  results  to  lower  transversal momenta of  light hadrons.  
Perhaps,   the inclusion of  such an effect into consideration  may  essentially 
improve not only our understanding  of   
the quark  coalescence process,
but also the formation of deuterons and other nuclear fragments in relativistic nuclear collisions.  
This subject is, however, outside the scope of the present work. 

As a typical VdW  EOS, the present model should be valid for  the low particle densities. 
Moreover, our analysis of the limit   $\mu / T \gg 1$ for fixed $T$  leads to  a surprisingly clear  kinetic expression 
for  the system's pressure \req{iiifourteen}.   Therefore, it is possible  that  this low density result
may provide 
a correct hint to study  the relativistic analog of the dense
packing problem.  
Thus,  it would be interesting to verify, whether the above approach remains valid  for relativistic quantum treatment  because  there  are  several unsolved problems  
for the  systems of  relativistic bosons and/or  fermions which, on one hand, are related to the problems discussed here and, on the other hand, 
may potentially  be important for relativistic nuclear collisions and   for  nuclear astrophysics.

\vspace*{0.9cm}

\noi
{\bf Acknowledgments.} The author thanks D. H. Rischke for the fruitful and  stimulating discussions,
and A. L. Blokhin for the important comments  on  the obtained  results. 
The research made in this work 
was supported in part   by the Program ``Fundamental Properties of Physical Systems 
under Extreme Conditions''  of the Bureau of the Section of Physics and Astronomy  of
the National Academy of Science of Ukraine. The partial  support  by the Alexander von Humboldt  Foundation is  greatly acknowledged.

\vspace*{0.5cm}

%%%%%%%%%%%%% Appendix A 

%\newpage

\bc
{\large \bf Appendix A: Relativistic Excluded Volume }
\ec

In order to study  the high pressure limit, it is necessary to 
estimate  the excluded volume of two ellipsoids, obtained by the Lorentz 
contraction of the spheres.  
In general, this is quite an involved problem. Fortunately, our analysis requires only the 
ultrarelativistic limit when the mean  energy per particle  is high  compared to  the mass of the particle.
The problem can be simplified further  since 
it is sufficient to find an analytical expression for 
the relativistic excluded volume with
 the  collinear particle velocities because the  configurations with 
 the  noncollinear  velocities
have larger excluded volume and, hence, are suppressed.
Therefore, one can safely consider the excluded volume   
produced  by two contracted cylinders (disks)  having the same
proper volumes as the ellipsoids.  
For this purpose the cylinder's height in the local rest frame is fixed to be   
$\frac{4}{3}$ of a  sphere radius.   

Let us introduce the  different radii $R_1$ and $R_2$ for  the cylinders, and 
consider for the moment  a zero height for the second cylinder $h_2 = 0$  and  
non-zero height $h_1 $ for  the first one.
Suppose that  the center of the coordinate system coincides with the geometrical 
center of the first cylinder  and  the  $OZ$-axis is perpendicular to the cylinder's  base.
Then the angle $\Theta_v$ between the  particle velocities is also  the angle between 
the  bases  of two cylinders.
To simplify the expression for the  pressure, the Lorentz frame is chosen to be the rest frame of the whole system. 

In order to estimate the excluded volume we fix the particle velocities 
and transfer the second cylinder around the first cylinder while  keeping the angle $\Theta_v$ 
fixed. The desired excluded volume is obtained as the volume 
occupied by the center of the
second cylinder under these transformations.
Considering the projection on the $XOY$ plane (see  Fig. 3.a),   
one should transfer the ellipse  with the semiaxes \mbox{$R_x = R_2 \cos \lp \Theta_v \rp$} 
and $R_y = R_2 $ around the circle of radius $R_1$. 
We approximate it by the circle of the averaged radius
$\bra R_{XOY} \ket = R_1 + (R_x + R_y)/2 = R_1 + R_2 ( 1 + \cos \lp \Theta_v \rp )/2$. 
Then the  first contribution to the excluded volume 
is the volume of the cylinder of the radius $\bra R_{XOY} \ket$ and 
the height  $h_1 = CC_1 $ of the cylinder $OABC$ in Figs. 3.a and 3.b, 
i.e.,   
\begin{equation}
v_{I} (h_1) = \pi \lp R_1 + \frac{R_2 ( 1 + \cos \lp \Theta_v \rp )}{2} \rp^2 h_1\,\,.
\end{equation}

Projecting the picture onto the $XOZ$ plane as it is shown in Fig. 3.b,
one finds that the translations of a  zero width disk   over the upper and lower 
bases of the first cylinder 
(the distance between the center of the disk and the base $CA$ is, 
evidently, $ CD_1 =  R_2 | \sin \lp \Theta_v \rp | $) 
generate the second conrtibution
to  the excluded volume
\begin{equation}
v_{II} (h_1) = \pi  R_1^2 \,2 \, R_2 | \sin \lp \Theta_v \rp | \,\,.
\end{equation}

The third  contribution follows from the translation of the disk  from  
the cylinder's base to the cylinder's side as it is shown for the
$YOZ$ plane in  Fig. 3.c.
The  area $BB_1F$ is the  part of
the ellipse segment whose magnitude  depends on the x coordinate. 
However, one can approximate it as the  quarter of the disk area  
projected onto the  $YOZ$ plane and can get a simple answer
$\pi  R_2^2 | \sin \lp \Theta_v \rp | / 4$. 
Since there are four of such transformations, and they apply for  
all x coordinates of the first cylinder (the length is $2\, R_1$), then 
the third contribution is
\begin{equation}
v_{III} (h_1) = \pi  R_1^2 \,2 \, R_1 | \sin \lp \Theta_v \rp | \,\,.
\end{equation}

%%%%%%%%%%%%%%%%%%%%%%%%%%%%%%%%%%%%

\begin{figure}\label{Cyl}
\mbox{\hspace*{1.0cm}\psfig{figure=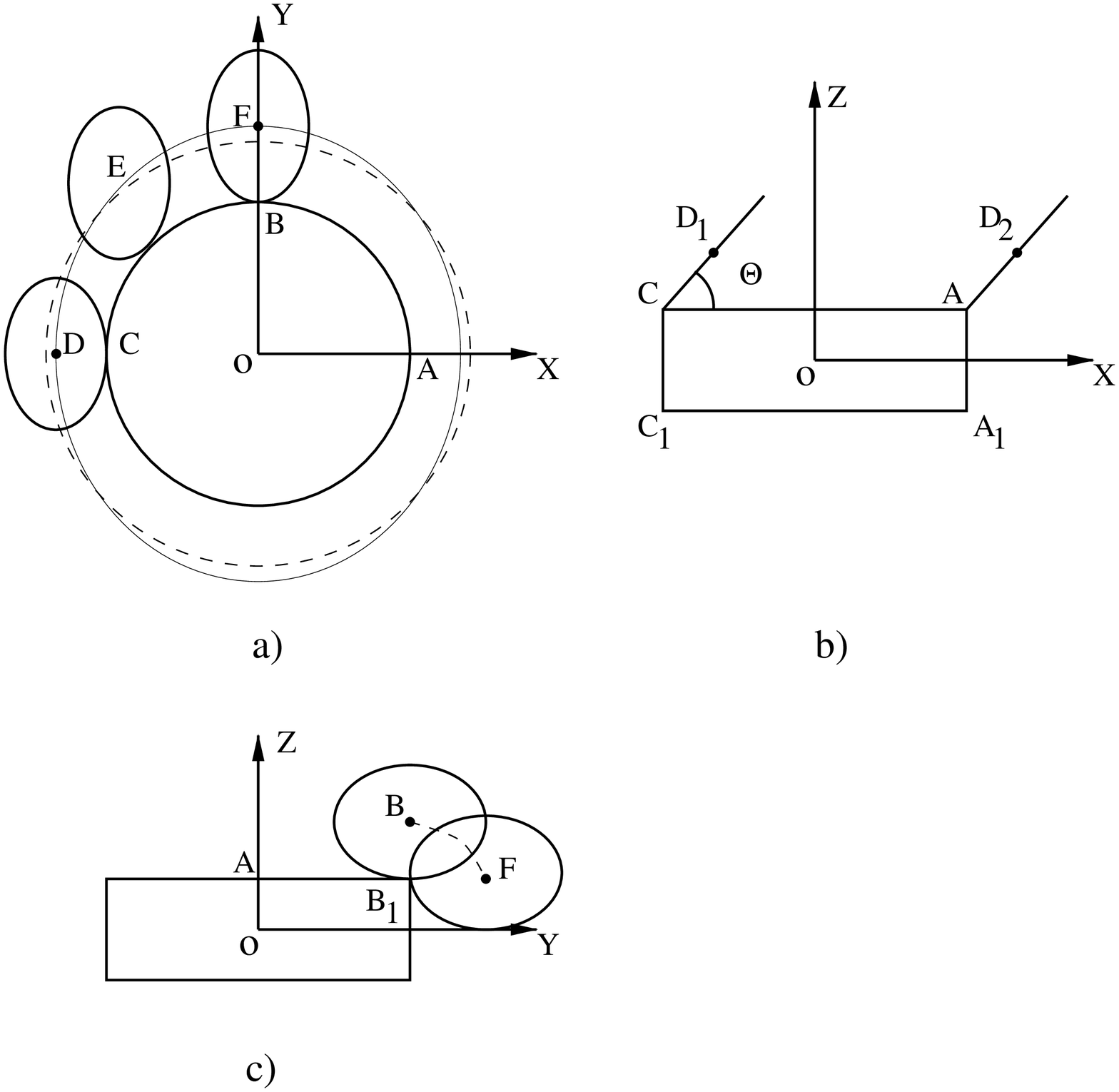,height=14cm,width=14cm}}\\

\vspace{0.25cm}

{\bf Fig. 3.} 
Relativistic  excluded volume derivation for relativistic 
cylinder $OABC$ and 
ultrarelativistic cylinder (disk)  $DC$ with radii $R_1$ and $R_2$, respectively.
$\Theta$ is the angle between their velocities.
Pictures a - c show the projections onto different planes.
The transfer of the cylinder $DC$  around the side of  
the cylinder $OABC$  is depicted in Fig. 3.a. 
The solid curve $DEF$ corresponds to the exact result, 
whereas  the dashed curve corresponds to the
average radius approximation
$\bra R_{XOY} \ket= OA + (DC + BF)/2 = R_1 + R_2 ( 1 + \cos \lp \Theta \rp)/2 $.
\\
%The corresponding approximative contribution to the excluded volume  
%is the volume of the cylinder with averaged radius $\bra R_{XOY} \ket$ and 
%the height $CC_1$ of the cylinder $OABC$.\\ 
%
The transfer of the cylinder $DC = DC_1 = AD_2$  
along  the upper base of the cylinder $OABC = ACC_1A_1$ 
is  shown in \mbox{panel b.} 
Its contribution to the excluded volume
is a volume of the cylinder with the base  $AC = 2 R_1$ and the height  
$CD_1 \sin \lp \Theta \rp = R_2 \sin \lp \Theta \rp $. 
A similar contribution corresponds to the disk transfer along 
the lower base of the cylinder $A_1C_1$.
\\
The third contribution to the relativistic excluded volume  
arises from the transformation of the  cylinder $DC = BB_1 = FB_1$  
from the upper base of the cylinder  $OABC = AB_1O$  to its side, and it
is schematically shown in Fig. 3.c.
The area $BB_1F \approx \pi / 4 R_2^2 \sin \lp \Theta \rp$  
is approximated as the one quarter of the area of the ellipse $BB_1$.
%Therefore, this part of the excluded volume is approximated as
%a volume of the cylinder with the square of the base equal to 
% $ \pi / 4 R_2 \sin \lp \Theta \rp$, and the height equal to $2 R_1$.
\end{figure}
%%%%%%%%%%%%%%%%%%%

Collecting all the contributions,  one obtains an estimate for the excluded volume of a cylinder and a disk 
\begin{equation}\label{vcuns}
v_{2c} (h_1) = \pi \lp  R_1 + R_2  \cos^2 \lp \frac{\Theta_v}{2} \rp \rp^2 h_1 + 
2 \, \pi  R_1 R_2 (R_1 + R_2) | \sin \lp \Theta_v \rp | \,\,.
\end{equation}

\noi
The above equation, evidently,  gives an exact result for a zero angle and arbitrary height of 
the first cylinder.
Comparing it with the exact answer for  $\Theta_v = \frac{\pi}{2}$
\begin{equation}\label{vcyldisk}
v_{2c}^{E} \lp h_1, \Theta_v = \frac{\pi}{2} \rp = 
R_1 \lp \pi  R_1 + 4\, R_2   \rp  h_1 +
2 \, \pi  R_1 R_2 (R_1 + R_2)  \,\,,
\end{equation}

\noi
one finds that the dominant terms (the second terms in \req{vcuns} and \req{vcyldisk})   
again are  exact, whereas the corresponding corrections, which are proportional to   $h_1$ ,
are related to each other  as
%%%$\frac{28,27}{28,56} \approx 0.9897$ 
$\approx 0.9897$ 
(ratio of the approximate to exact values at  $R_2 = R_1$).
Therefore, Eq. \req{vcuns} also gives a good approximation for the intermediate angles and 
small heights.

In order to get an expression for the non-zero height of the second 
cylinder we note that the expression  for the excluded volume   should be symmetric under the 
permutation of indexes 1 and 2. The lowest order correction 
in powers of the height  comes from the contribution $v_{I} (h_1)$. 
Adding the symmertic contribution $v_{I} (h_2)$ to  $v_{2c} (h_1) $ \req{vcuns}, 
one obtains the following result
\begin{eqnarray}
v_{2c} (h_1, h_2) & = & \pi \lp  R_1 + R_2  \cos^2 \lp \frac{\Theta_v}{2} \rp \rp^2 h_1 +
\pi \lp  R_2 + R_1  \cos^2 \lp \frac{\Theta_v}{2} \rp \rp^2 h_2  \nn
& + & 2 \, \pi  R_1 R_2 (R_1 + R_2)  \sin \lp \Theta_v \rp  \,\,.
\end{eqnarray}

\noi
The above expression  gives an exact result for a zero angle and arbitrary heights of cylinders.  
It also gives nearly exact answer for $\Theta_v = \frac{\pi}{2}$ in either limit
$h_1$ or $h_2 \rightarrow 0$.

Choosing the heights to reproduce the proper volume for  each of the Lorentz contracted 
spheres, one gets an approximation for the excluded volume of contracted spheres
in ultrarelativistic limit
\begin{eqnarray}
 v^{Urel}_{12}(R_1, R_2) & =  &  \frac{4 }{3 } \pi  \frac{ R_1}{ \g_1}  \lp  R_1 + 
R_2  \cos^2 \lp \frac{\Theta_v}{2} \rp \rp^2  +
\frac{4}{3}  \pi \frac{R_2}{\g_2} 
 \lp  R_2 + R_1  \cos^2 \lp \frac{\Theta_v}{2} \rp \rp^2  \nn
\label{vunnor}
& + & 2 \, \pi  R_1 R_2 (R_1 + R_2)  \sin \lp \Theta_v \rp  \,\,.
\end{eqnarray}
The  corresponding $\g_q$-factors ($ \g_q \equiv E(\1 k_q)/ m_q$,  $q = \{1, 2\}$) are defined  in the local rest frame of the whole system for particles of mass $m_q$.
The last result  is valid for $0 \le \Theta_v \le \frac{\pi}{2} $, to use it for 
$ \frac{\pi}{2} \le \Theta_v \le \pi $ one has to replace  
$ \Theta_v \longrightarrow \pi -  \Theta_v$ in (\ref{vunnor}). 

It is necessary  to stress  that the above formula gives a surprisingly  
good approximation even in nonrelativistic limit for 
the excluded volume of two spheres.
For $R_2 = R_1 \equiv R$ one finds that the  maximal excluded volume corresponds 
to the  angle $\Theta_v = \frac{\pi}{4}$ and its value is \mbox{$\max \{ v^{Urel}_{12} (R, R) \}
\approx \frac{36}{3} \pi R^3$},  whereas  the  exact result for nonrelativistic
spheres is $ v_{2s} = \frac{32}{3} \pi R^3$, i.e., 
the ultrarelativistic formula \req{vunnor}  describes a nonrelativistic
situation with the maximal deviation of about $10 \%$ (see the left panel in Fig. 4).

Eq. \req{vunnor} also describes the excluded volume $v_{sd} = \frac{10 + 3 \pi}{3} R^3$ for a nonrelativistic sphere and ultrarelativistic ellipsoid  with the maximal deviation 
from the exact result
of about $15 \%$ (see the right panel in Fig. 4).

%%%%%%%%%%%%%%%%%%%%%%%%%%%%%%%%%%%% Figures

%\clearpage

\begin{figure}[t]

\mbox{
\hspace*{0.0cm}\psfig{figure=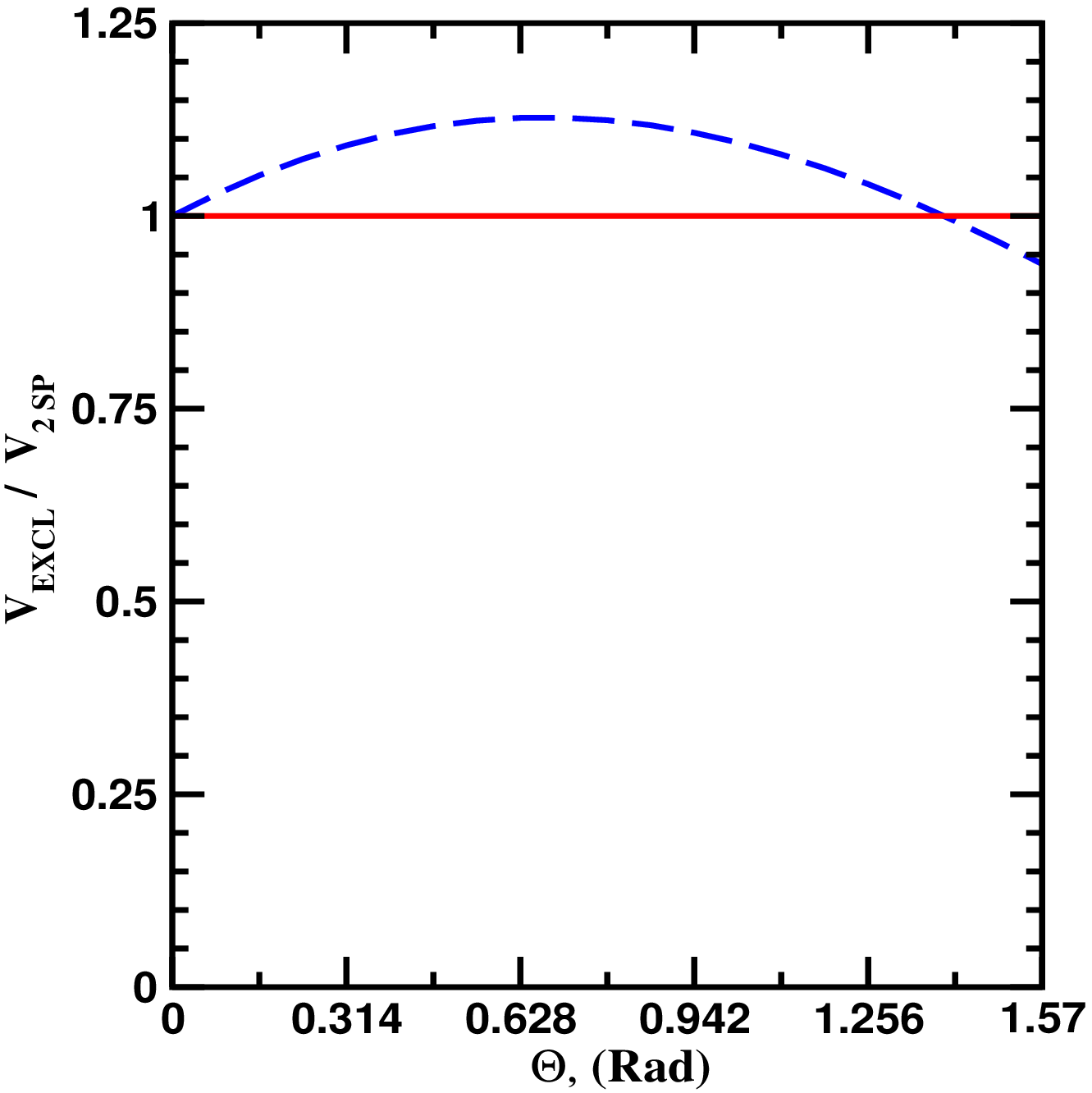,height=7.0cm,width=8.cm} 
\hspace*{-0.5cm} \psfig{figure=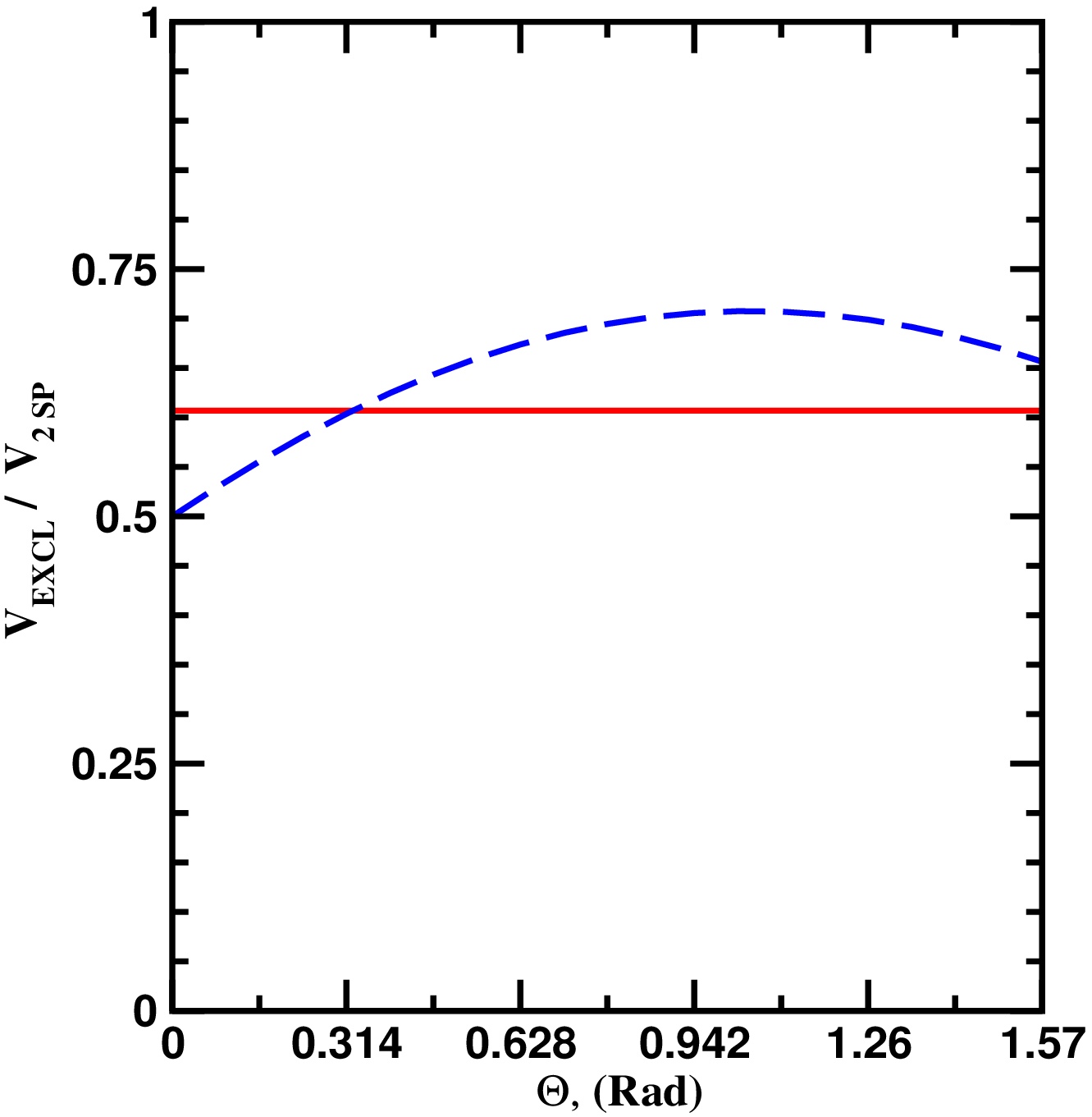,height=7.0cm,width=8.cm}
}

\vspace*{0.3cm}

{\bf Fig. 4.} 
Comparison of the relativistic  excluded volume obtained by the 
approximative ultrarelativistic formula 
with the exact results. 
The left panel shows the quality of the approximation $V_{EXCL} \equiv v^{Urel}_{12} (R, R) $ 
\req{vunnor} to 
describe the excluded volume 
of  two nonrelativistic spheres $V_{2 SP}$ of the same radius $R$ as  
a function of the spherical angle $\Theta$.
The right panel depicts the approximation to the excluded volume 
of the nonrelativistic sphere and disk. 
In both panels the solid curve  corresponds to the exact result and the long dashed one 
corresponds 
to the ultrarelativistic approximation by two cylinders.   
The averaged ultrarelativistic  excluded volume 
in the left panel is 
 $\frac{ \bra V_{EXCL} \ket_{\Theta} }{ V_{2 SP} } \approx 1.065 $.
The corresponding averaged value for the  right  panel 
is $\frac{ \bra V_{EXCL} \ket_{\Theta} }{ V_{2 SP} } \approx 0.655 $, which
should be compared with the exact value 
$\frac{ \bra V_{EXCL} \ket_{\Theta} }{ V_{2 SP} } \approx 0.607 $.

\end{figure}

In order to improve the  accuracy of (\ref{vunnor}) for  nonrelativistic
case, we introduce a factor $\a$ to normalize the integral
of the excluded volume \req{vunnor}
over the whole solid angle to the volume of two spheres
\begin{equation}\label{vcorr}
v^{Nrel} (R_1, R_2)
= \a \,\, v^{Urel}_{12} (R_1, R_2)\,; \hspace*{0.9cm}
\a = \frac{4  \p \lp  R_1 + R_2  \rp^3 }{\textstyle  3 \, \biggl. \int\limits_{\,0}^{\pi} 
d \Theta_v
\, \sin\lp \Theta_v \rp \,  v^{Urel}_{12} (R_1, R_2) \biggr|_{\g_1 = \g_2 = 1}
}
\,\,.
\end{equation}

For the equal values of hard core radii  and equal masses of particles  the normalization factor reduces to the following value
$\a \approx \frac{1}{1.0654}$, i.e., it compensates the  most of the deviations
discussed above.
With such a correction the excluded volume \req{vcorr}  can be safely used for
the nonrelativistic domain because in this case 
the VdW excluded volume effect is itself  a correction to the ideal gas
and, therefore, the remaining deviation from the exact result is of a higher order.

It is useful to have the relativistic excluded volume expressed
in terms of 3-momenta 
\begin{eqnarray}
v^{Urel}_{12} (R_1, R_2) & = &   \frac{ v_{01}}{ \g_1}  \lp 1 +
R_2   
\frac{ |\1 k_1| |\1 k_2| + |\1 k_1 \cdot \1 k_2| }{2 \, R_1\,|\1 k_1| |\1 k_2|} 
\rp^2  
+
\frac{ v_{02}}{ \g_1}  \lp 1 +
R_1   
\frac{ |\1 k_1| |\1 k_2| + |\1 k_1 \cdot \1 k_2| }{2 \, R_2\,|\1 k_1| |\1 k_2|} 
\rp^2 
\nn
& + & 2 \, \pi  R_1 R_2 (R_1 + R_2) 
\frac{ |\1 k_1 \times \1 k_2| }{|\1 k_1| |\1 k_2|}
\,\,,
\end{eqnarray}

\noi
where $v_{0q}$ denote the corresponding proper volumes 
$v_{0q} = \frac{4}{3} \pi R_q^3\,, \quad  q=\{ 1, 2\} $.  

For the practical calculations it is necessary to  
express the relativistic excluded volume in terms of the 
three 4-vectors - the two  4-momenta  of particles, $k_{q \, \mu}$,  and
the collective 4-velocity $ u^\mu = \frac{1}{\sqrt{1 - \1 v^2}} ( 1, \1 v)$.  
For this purpose one should reexpress the  gamma-factors and  at least one of    
 trigonometric  functions in \req{vunnor}  in a covariant form 
\begin{equation}\label{covar}
\g_q = \frac{\sqrt{m^2 + \1 k^2_q}}{m} = \frac{k_q^\m \, u_\m}{m}, \quad 
\cos\lp \Theta_v \rp = \frac{ k_1^\m \, u_\m \,\, k_2^\n \, u_\n - 
k_1^\m \, k_{2\,\m} }{
\sqrt{\lp (k_1^\m \, u_\m )^2 - m^2 \rp \lp (k_2^\m \, u_\m )^2 - m^2 \rp  } 
}\,\,.
\end{equation}
Using Eq. \req{covar},  one can express any trigonometric function of $ \Theta_v$ in a covariant form. 

%MY MARK

%%%%%%%%%%%%%%%%%%%%%%%%%%%%%%%%%%%%% REFs

% journals
\def\np#1{{ Nucl. Phys.} {\bf #1}}
\def\prl#1{{ Phys. Rev. Lett.} {\bf #1}}
\def\jp#1{{ J. of Phys.} {\bf #1}}
\def\zp#1{{ Z. Phys.} {\bf #1}}
\def\pl#1{{ Phys. Lett.} {\bf #1}}
\def\pr#1{{ Phys. Rev.} {\bf #1}}
\def\hip#1{{ Heavy Ion Physics} {\bf #1}}
\def\prep#1{{ Phys. Rep.} {\bf #1}}

%\newpage

%%%%%%%%%%%%%%%%
\end{document}